\documentclass[10pt,a4paper,abstracton,bottombeside]{scrartcl}
\usepackage{scrlayer-scrpage}
\addtokomafont{pagenumber}{\footnotesize}

\usepackage[T1]{fontenc}
\usepackage{amsmath}
\usepackage{xcolor}
\usepackage{authblk}
\usepackage{graphicx} \graphicspath{{./figures/}}
\usepackage[hidelinks]{hyperref}
\hypersetup{linkcolor=black,citecolor=black, urlcolor=black}
\usepackage[locale = US,separate-uncertainty = true,multi-part-units=single,range-units=single,list-units=single,detect-all]{siunitx} 
\usepackage[round,numbers,sort&compress]{natbib} 
\DeclareSIUnit{\arbitraryunit}{a.u.}
\usepackage{geometry}
\usepackage[font=scriptsize]{caption}
\usepackage{amsmath,amssymb,amsfonts}
\usepackage{times}
\usepackage{mathastext}

\usepackage{nicefrac} 
\usepackage{multirow,subcaption}
\usepackage{tcolorbox}
\definecolor{blau}{HTML}{7570b3}
\usepackage{booktabs,diagbox}
\usepackage{tikz}

\title{Complexity-based Encoded Information Quantification in Neurophysiological Recordings}

\author[a]{Julian~Fuhrer\footnote{Corresponding author e-mail: \href{mailto:julianpf@ifi.uio.no}{julianpf@uio.no}}$^{,}$}
\author[b]{Alejandro~Blenkmann}
\author[b,c]{Tor~Endestad}
\author[b,c,d]{Anne-Kristin Solbakk}
\author[a]{Kyrre~Glette}

\affil[a]{RITMO, Department of Informatics, University of Oslo, 0373 Oslo, Norway}
\affil[b]{RITMO, Department of Psychology, University of Oslo, 0373 Oslo, Norway}
\affil[c]{Department of Neurosurgery, Oslo University Hospital, Rikshospitalet, 0372 Oslo, Norway}
\affil[d]{Department of  Neuropsychology, Helgeland Hospital, 8657 Mosjøen, Norway}

\date{}
\setcapindent{0pt}

\newcounter{mysfig}[figure]
\renewcommand\themysfig{\alph{mysfig}}
\makeatletter
\newcommand\Scaption[1]{
	\refstepcounter{mysfig}
	\vskip.5\abovecaptionskip
	\sbox\@tempboxa{\textbf{\footnotesize(\themysfig)}~\footnotesize#1}%
	\ifdim \wd\@tempboxa >\hsize
	\textbf{\footnotesize(\themysfig)}~\footnotesize#1\par
	\else
	\global \@minipagefalse
	\hb@xt@\hsize{\hfil\box\@tempboxa\hfil}
	\fi
	\vskip\belowcaptionskip
}
\makeatother

\makeatletter 
\renewcommand\@biblabel[1]{#1.} 
\makeatother

\begin{document}
	
\maketitle
\lofoot[\copyright{} {\footnotesize 2022 IEEE. Personal use of this material is permitted. Permission from IEEE must be obtained for all other uses, in any current or future media, including reprinting/republishing this material for advertising or promotional purposes, creating new collective works, for resale or redistribution to servers or lists, or reuse of any copyrighted component of this work in other works.}]{}

\vspace{-15mm}
\begin{abstract}
	Brain activity differs vastly between sleep, cognitive tasks, and action. Information theory is an appropriate concept to analytically quantify these brain states. Based on neurophysiological recordings, this concept can handle complex data sets, is free of any requirements about the data structure, and can infer the present underlying brain mechanisms.
	Specifically, by utilizing algorithmic information theory, it is possible to estimate the absolute information contained in brain responses. 
	While current approaches that apply this theory to neurophysiological recordings can discriminate between different brain states, they are limited in directly quantifying the degree of similarity or encoded information between brain responses.
	Here, we propose a method grounded in algorithmic information theory that affords direct statements about responses' similarity by estimating the encoded information through a compression-based scheme.
	We validated this method by applying it to both synthetic and real neurophysiological data and compared its efficiency to the mutual information measure.
	This proposed procedure is especially suited for task paradigms contrasting different event types because it can precisely quantify the similarity of neuronal responses.
			\par\vskip\baselineskip\noindent
	\footnotesize\textit{EEG $|$ SEEG $|$ ECOG $|$ MEG $|$ Information Content $|$ Kolmogorov Complexity $|$ Entropy $|$ Lossless Compression}
\end{abstract}

\section{Introduction}
	A core task of the brain is to continuously encode, integrate, and store \textit{information} coming from the steady input stream of the sensory organs. Neurophysiological methods, such as electroencephalography~(EEG), intracranial EEG~(iEEG), magnetoencephalography~(MEG) or functional magnetic resonance imaging~(fMRI), are used to study the mechanisms the brain has evolved to accomplish this. 
	Given the analogy between these mechanisms and principles from information theory \citep{Timme18}, it appears plausible to employ ideas from the latter to analyze neuroimaging data. This mathematical theory provides multivariate analysis tools, is not bound to a single type of data, is model-independent (i.e., it does not require assumptions about the data itself) and can capture nonlinear interactions \citep{Timme18, Li08, Zbili21}. 
	These properties make information theory an apt approach to effectively treat neuroimaging data. 
	In particular, the discipline of \textit{algorithmic} information theory~(AIT) is especially suited because it can obtain an absolute value of information for individual responses. For example, it has been utilized to analyze cognitive operations or to discriminate between states of consciousness measured with EEG, iEEG, MEG or fMRI recordings~\citep{Sitt14, Fuhrer21, Canales-Johnson20, Schartner17, Schartner15, Varley20}. 
	However, the applied methods are limited in their possibility to directly quantify the level of similarity or encoded information between response. Therefore, studies often use measures that draw on the concept of Shannon entropy (i.e., classic information theory), such as the popular measure of mutual information.
	
	\begin{figure}[htbp]
		\centering
		\smallskip\smallskip
		\includegraphics[width=.5\linewidth]{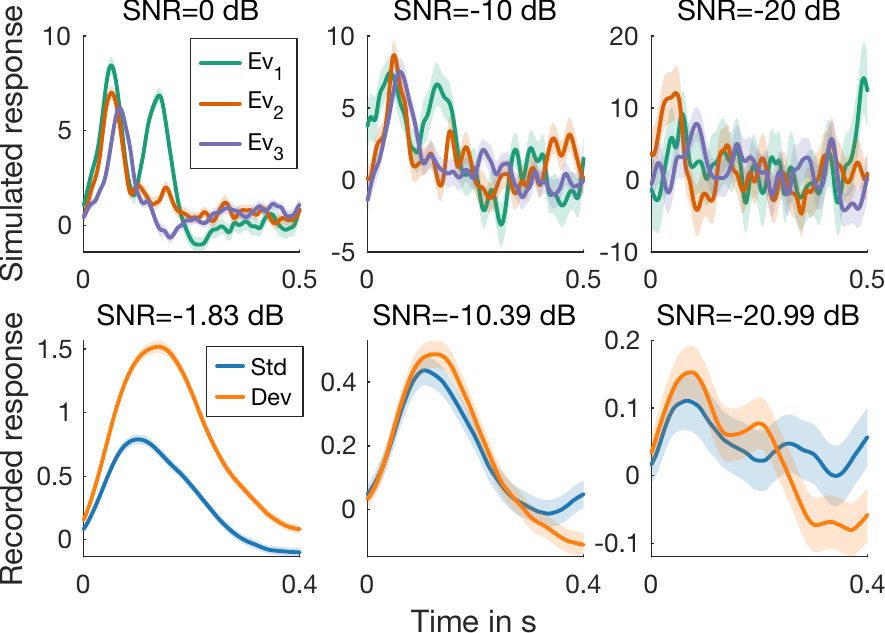}
		\caption{Example signals. Top: Event related potentials for each event type simulated with the neural mass model for different SNRs (model input function for Ev$_{1}$: Boxcar, for Ev$_2$: Exponential decay, and for Ev$_3$: Mexican hat). Bottom: Example evoked HFA responses for standard and deviant tones from the oddball paradigm data possessing similar SNRs.}
		\label{fig:pltoutputs}
	\end{figure}
	
	This paper proposes an approach to quantify neurophysiological recordings through information-theoretical principles. 
	We investigated a universal distance measure grounded in AIT that allows to directly estimate the similarity between individual event types. 
	Named Normalized Compression Distance~(NCD), it computes a normalized similarity measure given two signals' absolute information content estimates~\citep{Li04, Li08}. This measure has been applied to, for example, cluster analysis, virology, language, music, and literature \citep{Li08}, but to our knowledge it has only been used once in the field of neuroscience \citep[i.e., for brain-computer interfaces,][]{Sarasa19}. 
	In the remainder of this paper, we propose an approach for applying this measure to quantify differences in neurophysiological data, which is especially suited for oddball task paradigms. We validated its use through both a synthetic and an experimental scenario, where the latter emerged from an auditory oddball paradigm with iEEG recordings \citep{Blenkmann19}. We demonstrate its performance on signals with varying signal-to-noise-ratio~(SNR) and compare its efficiency to that of the mutual information measure~(MI). 


\renewcommand*{\thefootnote}{\fnsymbol{footnote}}\setcounter{footnote}{1}
\section{Materials \& Methods}

\subsection{Electrical brain activity} \label{sec:EEG}
Two types of experimental scenarios were assessed. In the first scenario, we examined information processing parameters by simulating electrical brain activity through the neural mass model from \citet{Jansen95}. This model has been widely used, including the simulation of intracranial brain activity \citep{Jacobacci13, Wendling00}. We simulated two independent cortical columns through this model. One column acted as an task-evoked node and the other as a pure noise-producing node. We used three different functions as input to the evoked node to mimic the onset of possible presentations of different event types: A Boxcar function, an exponential decay, and a Mexican Hat function. All inputs had a maximum amplitude of \SI{100}{\arbitraryunit} and produced the characteristic evoked responses required for the evoked node. Lower amplitudes lead to undesired noisy signals. The noise node had a constant input of \SI{100}{\arbitraryunit}. To excite the model, Gaussian white noise was subsequently added to all input signals. 
We imitated real EEG recordings, by simulating each event type 50 times, constituting 50 trials (event responses). Each trial was \SI{500}{\milli\second} long (501~samples). Both nodes were combined by demeaning the trials from the noise node and subsequently taking the average of both nodes on a trial-by-trial basis~(Fig.~\ref{fig:pltoutputs}).

In the second scenario, we evaluated our method on real experimental iEEG data obtained from intracranial electrodes implanted in normal-hearing adults with drug-resistant epilepsy. Participants (n=22) performed a passive auditory oddball paradigm where a standard tone alternated with random deviant tones \citep[for details, see][]{Blenkmann19}. High-frequency activity~(HFA; \SIrange{75}{145}{\hertz}) was extracted, and differences between standard and deviant tone responses were evaluated in the \SI{400}{\milli\second} time window (401~samples) following the sound onset across channels and subjects~(Fig.~\ref{fig:pltoutputs}).

\begin{figure*}[htbp]
	\centering
	\smallskip\smallskip
	\includegraphics[scale=1]{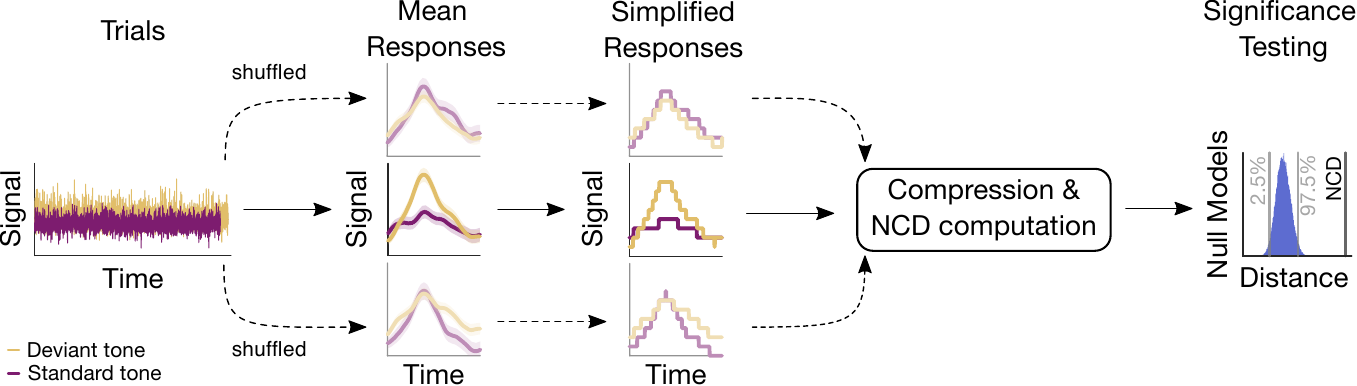}
	\caption{Sketch of the setup: Mean responses were computed for each test condition (here: standard tones and tones deviating from those), given the recorded trials of an experiment. These mean responses were then simplified and underwent compression to estimate their complexity and NCD values. We then assessed the significance of the NCD through surrogate data testing: Trials were randomly shuffled across conditions. Based on that, surrogate mean responses (dashed arrows) were computed, simplified and subsequently compressed. This resulted in a null model distribution for the NCD.}
	\label{fig:setup}
\end{figure*}

\subsection{Mean Response}
EEG recordings reflect the activity of several ongoing processes in the brain. Consequently, EEG captures both related and unrelated activity for a process of interest. Activity that is relevant to a specific process is typically obtained by computing a mean response that averages many trials such that irrelevant brain activity averages out. In this vein, based on a given signal~$z\in\mathbb{R}^{\mathrm{T}\times\mathrm{N}}$ with T~trials and N~samples from one channel, we identify the mean response of the channel as $\bar{z}\in\mathbb{R}^\mathrm{N}$ (Fig.~\ref{fig:pltoutputs}\&\ref{fig:setup}).

\subsection{Signal-to-noise Ratio}
To evaluate our method in terms of noise within signals, we used the SNR measure defined as 
\begin{equation}
	\mathrm{SNR}=10\log\frac{\sum_{\mathrm{k}=1}^{\mathrm{N}}u^2_\mathrm{k}}{\sum_{\mathrm{k}=1}^{\mathrm{N}}n^2_\mathrm{k}} \,,
	\label{eq:SNRa}
\end{equation}
with signal length~N, a signal of interest~$u\in\mathbb{R}^\mathrm{N}$, and noise~$n\in\mathbb{R}^\mathrm{N}$. In the synthetic scenario, we used this definition to weigh the output from the noise node to attain desired SNRs. In the scenario with direct brain recordings, SNRs are approximated from the data, where the noise component is inferred by subtracting the mean signal~$\bar{z}$ from a recorded signal~$z\in\nobreak\mathbb{R}^{\mathrm{T}\times\mathrm{N}}$. This leads to
\begin{equation}
	\mathrm{SNR}=10\log\frac{\sum_{\mathrm{k}=1}^{\mathrm{N}}\bar{z}^2_\mathrm{k}}{\sum_{\mathrm{k}=1}^{\mathrm{N}}\sigma^2_\mathrm{k}} \,,
	\label{eq:SNRb}
\end{equation}
where $\sigma\in\mathbb{R}^\mathrm{N}$ is the variance of the recorded signal (i.e., an estimate of the noise variance). When applying Eq.~\ref{eq:SNRa}\&\ref{eq:SNRb} to the synthetic data, the SNRs coincide.

\subsection{Information Content Estimation}
Given a mean response, we estimated its information content by employing a concept of AIT. This theory draws on the idea of algorithmic complexity or Kolmogorov Complexity (K-complexity). The K-complexity is the ultimate compressed version or minimum description length of an object, i.e., its absolute information content \citep{Li08}. If the minimum description length is short (long), an object is characterized as "simple" ("complex"). Because it is not possible to compute the theoretically ideal K-complexity, it is often heuristically estimated, obtaining an upper-bound approximation. Possible estimation approaches are conventional lossless data compression programs, e.g., gzip \citep{Li08, Li04}.
Based on the K-complexity, various metrics were derived. The theoretical Normalized Information Distance~(NID) is one such instance. For a pair of signals~$(x,y)$ it is defined as
\begin{equation*}
	\mathrm{NID}(x,y) =  \frac{\max(K(x|y),K(y|x))}{\max(K(x), K(y))} \,,		
\end{equation*}
where $K(x|y)$ is the K-complexity of x given y. It allows to compare different pairs of objects with each other and suggests similarity based on their dominating features (or a mixture of sub-features) \citep{Li04, Li08}. For a pair of strings~$(x,y)$, its estimation counterpart, the NCD, is defined as 
\begin{equation}
	\mathrm{NCD}(x,y) = \frac{C(xy)-\min(C(x),C(y))}{\max(C(x),C(y))}\,,
	\label{eq:NCD}
\end{equation}
with $C(xy)$ denoting the compressed size of the concatenation of $x$ and $y$, and $C(x)$ and $C(y)$ their respective size after compression \citep{Li04, Li08}. Further, the NCD is non-negative, where smaller NCD values suggest similar, and higher values different objects.
We compared the NCD measure for EEG data to the vastly applied MI measure. This measure is also grounded in information theory. In contrast to the NCD, it draws on the concept of Shannon entropy (i.e., classic information theory). For a discrete random variable x with N outcomes, the entropy can be defined as
\begin{equation}
	H(x) = -\sum_{k=1}^\mathrm{N} {\mathrm{p}(x_\mathrm{k}) \log \mathrm{p}(x_\mathrm{k})}\,,
	\label{eq:H}
\end{equation}
with $\mathrm{p}(x_\mathrm{k})$ being the occurrence probability for each element~$x_\mathrm{k},\dots,x_\mathrm{N}$ of $x$. Given this definition, the MI between two discrete random variables~$(x,y)$ with N or M outcomes can be defined as 

\begin{align}
	\label{eq:I}
	\begin{split}
		\mathrm{MI}(x; y) &= H(x) - H(x \vert y) \\
		&= H(y) - H(y \vert x)\\
		&=\sum_\mathrm{k=1}^\mathrm{N} \sum_\mathrm{j=1}^\mathrm{M} p(x_\mathrm{k},y_\mathrm{j}) \log \frac{p(x_\mathrm{k},y_\mathrm{j})}{p(x_\mathrm{k})p(y_\mathrm{j})} \,,
	\end{split}
\end{align}
with the joint probability~$p(x_\mathrm{k},y_\mathrm{j})$ and the marginal probabilities~$p(x_\mathrm{k})$ and $p(y_\mathrm{j})$. 

In our analysis, a signal was simplified (Fig.~\ref{fig:setup}) by grouping its values into discrete steps~(bins). The bins covered equal distances and in a range between the global extrema of all the signals considered. For the compression-based scheme, the compressor received the indices of the bins that contained the elements of the signal \citep{Sitt14, Canales-Johnson20}. Compression then proceeded through a compression routine based on Python’s standard library with gzip\footnote{Custom analysis codes written in Matlab are available at \href{https://osf.io/tnvc4}{osf.io/tnvc4}}.

\subsection{Significance Testing}
To estimate the statistical significance of the information-based measures, we utilized surrogate data testing \citep{Lancaster18}. Accordingly, p-values were obtained by evaluating the observed information-based quantity regarding a null distribution (Fig.~\ref{fig:setup}). Null distributions were created by repeatedly shuffling the trials (i.e., single evoked responses) between conditions (e.g. standard and deviant) and then re-computing the information-based measure. Single sided p-values lower or equal to 0.05 were considered statistically significant.

\subsection{Detection Rate}
We assessed the performance of our measure using a detection rate, i.e., the ratio between significant p-values and the total amount of p-values. For the simulation-based scenario, a simulation experiment was repeatedly realized, yielding a distribution of p-values. The simulation experiment employed three distinct input types (see Sec. ~\ref{sec:EEG}), producing clear differences in their outputs (Fig.~\ref{fig:pltoutputs}). We then computed a mean detection rate based on the pairwise comparison of all inputs to obtain a robust measure. 
For the experimental scenario, we determined detection rates from the p-value distributions of channels grouped by similar SNRs.


\section{Results}
\subsection*{Scenario~I: Simulation Experiment}
We evaluated our method’s performance by systematically manipulating the signals’ SNR from \SIrange{-20}{20}{\decibel}. Thus, we assessed how it operates on a range of noise levels, which allows us generalization of its performance to other data sets. Besides the signals’ SNR, we simultaneously varied the number of bins from \numrange{2}{128}, thereby determining the level on which signal values are grouped. We evaluated to which extent the detection rate depends on the number of bins given different SNRs.
Our results exhibited a clear distinction between low and high detection rates for the NCD, which was not apparent for the MI (Fig.~\ref{fig:res}\ref{fig:sDR}). These results emerged from 100~simulation experiment realizations and \num{1e3}~null models for each SNR and bin combination (i.e. \num{5.17e6} realizations in total). In addition, the mean detection rate was \SI{7.03}{\percent} higher for NCD across all SNRs and number of bins (\SI{61.08}{\percent} vs. \SI{54.05}{\percent}, Tab.~\ref{ta:mDR}, Mann–Whitney–Wilcoxon tests with p\num{<1e-106} for all pairs). On average, the NCD-based method shows higher rates than the MI-based method, expect for very low number of bins.

\begin{table}[htbp]
	\centering
	\caption{Mean Detection Rates in \SI{}{\percent}.}
	\begin{tabular}{cccc}
		\diagbox[]{SNR}{Bins}& \numrange{2}{128} &  $\num{<32}$ & $\num{\geq32}$\\ 
		\toprule
		& NCD\,\, MI & NCD\,\, MI & NCD\,\, MI\\
		\midrule
		$\SIrange[]{-20}{20}{\decibel}$ & 61.08\,\, 54.05 & 45.46 \,\, 63.17 & 65.98\,\, 51.47\\ 
		$\SI{>0}{\decibel}$ 		    & 88.72\,\, 79.31 &  65.31 \,\, 86.68 & 95.95\,\, 77.04\\
		$\SI{\leq 0}{\decibel}$			& 34.76\,\, 30.00 &  26.54 \,\, 40.78 & 37.30\,\, 26.66\\						
		\bottomrule
	\end{tabular}
	\label{ta:mDR}
\end{table}

\subsection*{Scenario~II: Direct Brain Recordings}
\begin{figure*}[htbp]
	\centering
	\smallskip
	\begin{minipage}[t]{.28\textwidth}
		\begin{minipage}[b]{\columnwidth}
			\Scaption{Mean Detection Rates for NCD\label{fig:sDR}}\vspace{2.5mm}
			\includegraphics[scale=.85]{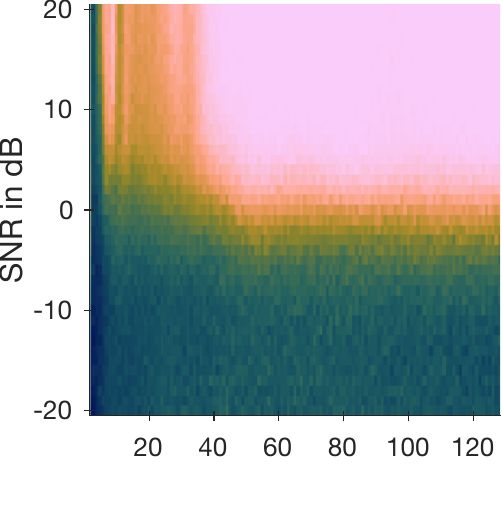}
		\end{minipage}
	\end{minipage}
	\begin{minipage}[t]{.3\textwidth}
		\begin{minipage}[b]{\columnwidth}
			\Scaption{Mean Detection Rates for MI\label{fig:sDR2}}\vspace{2.5mm}
			\includegraphics[scale=.85]{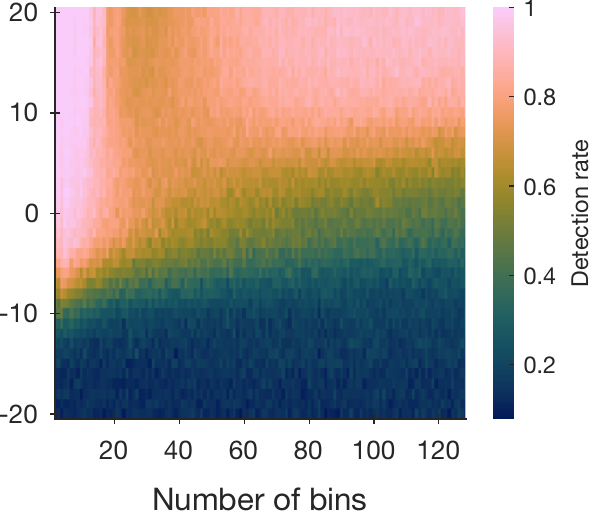}
		\end{minipage}
	\end{minipage}
	\hspace{6mm}
	\begin{minipage}[t]{.3\textwidth}
		\begin{minipage}[b]{\columnwidth}
			\Scaption{Difference in Detection Rates\label{fig:sDR3}}\vspace{2.5mm}
			\includegraphics[scale=.85]{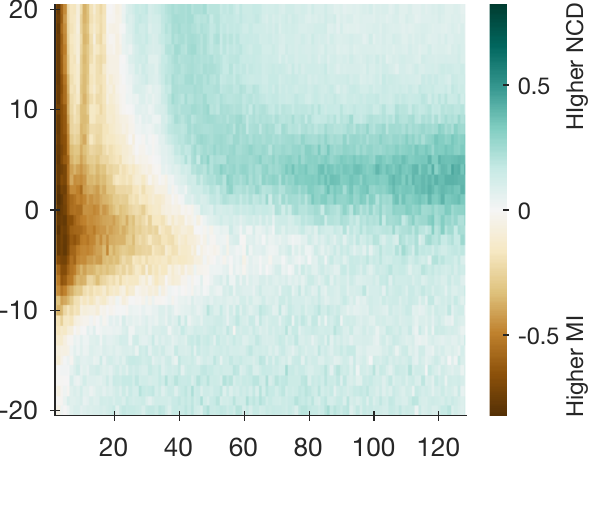}
		\end{minipage}
	\end{minipage}
	\Scaption{Detection rates over SNR for different number of bins \label{fig:eDR}}\vspace{2.5mm}
	\includegraphics[scale=.9]{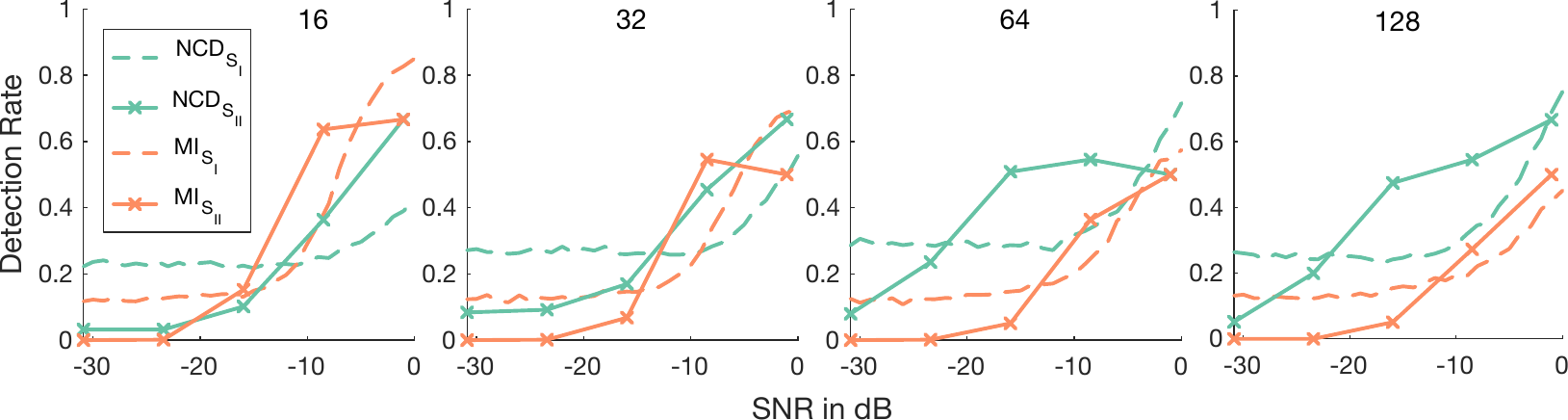}
	\caption{\textbf{a}:~Mean detection rate across number of bins and SNRs for NCD. Depicted is the detection rate averaged across the three event types, stemming from 100 realizations of the simulation experiment for each SNR to bin combination. 
		\textbf{b}:~The same results but for MI. 
		\textbf{c}:~Difference in the detection rates between NCD and MI.
		\textbf{d}:~Detection rate emerging from the HFA data set. To account for the uneven distribution of SNR values, channels within a SNR interval of \SI{6}{\decibel} were aggregated. For comparison dashed lines are added, originating from \num{1e3} realizations of the experiment in scenario~I.}
	\label{fig:res}
\end{figure*}

Given evoked HFA responses from a total of 1078 iEEG channels pulled from 22 subjects, we investigated how our method discriminates between standard and deviant tones. For each channel, we estimated its SNR and applied our approach given standard or deviant tone responses. The mean number of trials per subjects was \num{759.50+-360.85} for deviant responses and \num{715.14+-388.12} for standards. To account for the unevenly distributed SNRs across channels, we binned the data into \SI{6}{\decibel} SNR bins. That allowed us to compare the results with scenario~I. As visualized in Fig.~\ref{fig:res}\ref{fig:eDR}, detection rates were similar to those of scenario~I, showing on average higher detection rates for the NCD~measure, in particular for the higher number of bins.

\section{Discussion}
We proposed a procedure grounded in information theory to assess differences between brain responses in electrophysiological data while vastly reducing their dimension. By applying this procedure to two scenarios of EEG recordings, we showed that it effectively discriminates between responses to different event types. Moreover, our method turns out to be more sensitive than the popular measure of mutual information.

The basis of our scheme is the computation of a mean response per event, given repeatedly measured trials. Alternatively, it is also possible to directly compress the concatenation of trials (Fig.~\ref{fig:setup}). However, because each trial is inherently noisy, more trials lead to more noise being contained within the concatenation. That causes a decrease in the compressor effectiveness merely based on the number of trials. On the other hand, by determining a mean response from all trials, the noise contained across trials potentially cancels out. That enables a more effective compression while the shuffling of trials between event types during significance testing still accounts for their variance. 
Relevant to consider are potential sources where the noise originates from. Researchers \citep{Echeveste20} still debate the role of the so-called noise variability, which could be especially relevant for intracranial EEG recordings. Also, the employed compressor might play a role. Modern compressors such as Brotli~\citep{Alakuijala18} or LZMA~\citep{Pavlov22} might be better suited for compressing longer sequences. This remains to be investigated.

Another point worth highlighting is the general difference between NCD and MI during significance testing. High NCDs indicate different, while high MIs indicate similar signals. Accordingly, in the null distributions during significance testing (Fig.~\ref{fig:setup}), the observed NCD is found on the superior side of respective null distributions and MI on the inferior side. Shuffling trials mixes both event types, such that, relative to the measured quantity, the means of the null distributions tend to decrease for the NCD, while they tend to increase for the MI measure.

When increasing the number of bins (\raisebox{1.5pt}{\scalebox{.7}{$\gtrsim$}}\,32), results were more robust in the case of NCD (Fig.~\ref{fig:res}, Tab.~\ref{ta:mDR}). Accordingly, the opposite effect appears to occur for the reference quantity: The MI~detection rate rises the lower the number of bins -- for both scenarios (Fig.~\ref{fig:res}\ref{fig:sDR3}\&\ref{fig:eDR}). 
This is due to the consequent simplification of the signals (Fig.~\ref{fig:setup}), where a lower number of bins leads to fewer different probabilities during the MI~computation (Eq.~\ref{eq:H}\&\ref{eq:I}).
Accordingly, the surrogate entropy values of both conditions draw nearer to each other, such that the observed MI tends to be always situated outside of (i.e., lower than) the surrogate null distribution. For lower SNRs, again, the signals have a similar entropy to begin with. Consequently, respective null model distributions always contain observed MIs. On the other hand, the NCD measure does not exhibit an equal effect for lower numbers of bins, because the input size to the compressor is invariant to the number of bins.

It is important to point out that no universal procedure for the computation of SNRs exists, and SNRs are usually not reported for EEG recordings. For this reason, we described an easy and applicable procedure to estimate the SNR, making it possible to use our results (Fig.~\ref{fig:res}) for orientation. For high SNR values, the NCD provides the better results. This tendency continues for lower SNRs, yet both measures in common is the transition around \SI{0}{\decibel} (Fig.~\ref{fig:res}\ref{fig:sDR}\&\ref{fig:sDR2}).

A limitation in scenario~II is the predetermined unequal distribution of SNRs over channels. That caused us to group channels according to their SNR irrespective of their location in the brain. Since we were dealing with an auditory oddball paradigm and comparing responses to expected and unexpected tones, some brain regions might be more sensitive to the differences between sound types than others. Consequently, detection rates shown in Fig.~\ref{fig:res}\ref{fig:eDR} might be higher for regions more sensitive to sounds. This could be more easily investigated in scalp EEG recordings with a homogeneous distribution of channels. Further, the detection rates in scenario~I depend on the number of simulation experiment realizations. However, increasing the realizations from 15 to 100 only caused minor changes in detection rate.

\section{Conclusion}
We have presented an NCD-based method to compare neurophysiological recordings and applied it to synthetic and real EEG signals. 
Overall, our NCD-based method performs better than the MI-based alternative measure. The most relevant parameter behind this result is the number of bins employed during the binning procedure. In the simulation study, a higher number of bins gives more robust detection rates for the NCD, whereas for MI, too high densities cause a decline in detection rate. 
Common to both measures are low detection rates for low SNRs. However, the NCD performs slightly better on direct brain recordings from the oddball paradigm study. 
Future studies could explore this further through a hybrid NCD measure, i.e., combining AIT with classic information theory. Another possibility is to focus on the compression procedure by employing modern compressors, e.g., neural network compressors.
Taken together, our procedure grounded in AIT is an apt candidate to represent and investigate neurophysiological data through information-theoretical principles.

\bibliographystyle{IEEEtranN}
{\footnotesize
	\bibliography{refs}
}

\end{document}